\documentclass[12pt]{article}

\usepackage{amssymb,amsmath}

\begin{document}   
\newtheorem{Theorem}{Theorem}
\newcommand {\TR} {\text{TR}\,} 
\newcommand {\tr} {\text{tr}}
\newcommand {\gm} {\mathbf{g}}
\newcommand {\la} {\lambda}

%


\title{Star products and central extensions}

\author{Jouko Mickelsson\\
Department of Mathematics \\
University of Helsinki \\
and Mathematical Physics\\ 
Royal Institute of Technology, Stockholm}

\maketitle
 





{\bf Abstract}

The purpose of the present note is two-fold. First, to show that
deformations of algebras of smooth functions can be used to construct
topologically nontrivial standard
central extensions of loop groups. Second, to use noncommutative
geometry as a regularization of current algebras in higher dimensions
with the aim of constructing representations of current algebras. \newline\newline
\newline\newline
Dedicated to Krzystof P. Wojciechowski on his 50th
  birthday

\section{Introduction}\label{intsec}

The standard central extension of the algebra $L\gm$ of smooth loops 
in a Lie algebra $\gm$ of a compact Lie group $G$ defines a central
extension by the circle of the smooth loop group  $LG.$ An explicit
geometric construction for the central extension $\widehat{LG}$ was
given by Mickelsson  \cite{Mi}; for an alternative construction see Murray  
\cite{Mu}. 
The method in \cite{Mi} was to first define a topologically trivial central 
extension of the group $DG$ of smoooth $G$ valued functions in the 
unit disk $D$ and then take a quotient by a normal subgroup isomorphic 
with the group $\mathcal {G}$ of functions which take the value $1\in G$ on
the boundary circle. The central extension of $DG$ is defined by a
$S^1$ valued 2-cocycle. In Section \ref{disksec} we shall see that we can dispens
the 2-cocycle if we use a Moyal product for the functions in the
disk. The structure of the loop group on the boundary circle remains 
undeformed but we need a derminant in $\mathcal {G}$ associated to a trace 
functional on the Moyal algebra. 

The second application of the use of Moyal product for function
algebras is related to the problem of constructing nontrivial 
representations of current algebras arising from hamiltonian
anomalies, \cite{Mi2}. The main difficulty comes from the missing 
Hilbert-Schmidt property of off-diagonal elements of the currents 
with respect to the energy polarization. This problem does not arise in the 
case of current algebras on the circle (the lowest energy
representations are the highest weight representations of affine Lie 
algebras). However, in any dimension bigger than one  the
Hilbert-Schmidt condition fails; this is related to ultraviolet 
divergencies in perturbative Yang-Mills theory. In one space
dimension the divergencies  can be removed by normal ordering but 
in higher dimensions one needs additional subtractions. The (background field 
dependent) subtractions form an obstruction for constructing true Hilbert 
space representations; the best what one can achieve is a geometric 
action on sections of a Hilbert bundle over the space of background fields. 

A deformation of the commutative algebra of smooth functions on a
manifold can improve the short distance behaviour in quantum field
theory. One of the examples is the fuzzy sphere which has been studied 
in great detail by Grosse and Madore, \cite{GM}, \cite{Ma}. 
In this case the algebra becomes
finite-dimensional, avoiding any kind of ultraviolet divergencies.
Consequencies for the current algebra representations are illustrated 
in terms of three examples in Section \ref{gensec}. 

The algebra of functions on the disk can be deformed in a variety of
ways. A different construction can be found in an article by Lizzi, Vitale, and Zampini \cite{LVZ} which is more
close in spirit to the fuzzy sphere algebra in \cite{GM}, \cite{Ma}.

\section{The disk algebra and central extensions of loop groups}\label{disksec}

Let $\omega$ be the standard symplectic form $\omega= dx\wedge dy$ in $\mathbb R^2.$ 
Its restriction to the unit disk $D$ in $\mathbb R^2$ can be used to
define a star product deformation of the algebra 
$\mathcal {B}$ of complex $n\times n$ matrix valued smooth functions in 
$D$, with vanishing normal derivatives to all orders at the boundary $S^1$, 

\begin{eqnarray}
(f*g)(x,y) = e^{\frac{i\nu}{2}(- \partial_x\partial_{y'} + \partial_y\partial_{x'})} f(x,y)g(x',y')|_{x=x',y=y'},\label{eq1}
\end{eqnarray}
defined as a formal power series in $\nu.$ Note that at the boundary the star product is just the 
pointwise product of functions. Thus the restriction to the boundary
gives the trivial formal deformation of the loop algebra. For general background on Moyal product and 
deformation quantization see  Bayen, Flato, Fronsdal, Lichnerowicz, and Sternheimer 
\cite{BFFLS}.  

Integration over the disk defines a linear functional in $\mathcal {B}$, 
\begin{eqnarray}
\text{TR}_{\nu} (f) = \frac{1}{2\pi \nu} \int_D \text{tr}\,f(x,y) dx dy,\label{eq2}
\end{eqnarray}
where $'\text{tr}'$ is the matrix trace.

If the functions $f,g$ are constant  on the boundary then by integration by parts one observes that
\begin{eqnarray}
\text{TR}_{\nu}(f*g - g*f) =0.\label{eq3}
\end{eqnarray}
Otherwise, one has
\begin{eqnarray}
\text{TR}_{\nu}(f*g-g*f) =  \frac{1}{4\pi i} \int_D \text{tr}
\,(df dg-dgdf)+\dots =   
\frac{1}{2\pi i} \int_{S^1} \text{tr} \,f dg,\label{eq4}
\end{eqnarray}
where the dots denote terms containing higher derivatives in the
radial direction which integrate to zero through integration by parts due
to the boundary conditions. 
Thus $\text{TR}_{\nu}$ is a true trace only  in the subalgebra $\mathcal B_0 $ of functions 
constant on the boundary. We shall also use the complex trace 'TR' defined
as the zeroth order term in the formal Laurent series $\text{TR}_{\nu}.$  
This is
likewise a true trace on the algebra of functions vanishing at the
boundary. 

Any multiple of (\ref{eq2}) by a Laurent series in $\nu$ is also a trace 
on the star subalgebra of constant functions on the boundary. 
However, the choice of the normalization will become apparent later.
Actually, any
trace is proportional, up to a factor in $\mathbb C[\nu^{-1}, \nu]],$ to
the trace above,  Fedosov \cite{F}. (For a short proof in the manifold case see Gutt and Rawnsley \cite{GR}.) 

Let $G$ be a compact matrix group and $DG\subset \mathcal B$  be the group
of $n\times n$ matrix valued 
functions on $D,$ formal power series in $\nu,$ which are invertible with respect to the star
product and matrix multiplication and such that the boundary values belong to the
matrix group $G.$   Note that an inverse exists if and only if the zeroth
order term in $\nu$ is invertible as an ordinary matrix valued
function. 

The group $DG$ factorizes
to a product of two spaces. The first factor is the set $D_0G$ of zero order
functions in $DG$ and the second factor is the group $K$ of functions of
the form 
$$f= 1 + \nu f_1 +\nu^2 f_2 + \dots.
$$ 
Note that any $f$ of this type has an inverse as a formal power series
in $\nu.$ The group $K$ is contractible and it has a uniquely defined
logarithmic function taking values in the formal power series without
constant term.

We denote by $\mathcal G$ the subgroup of $DG$ consisting of functions
which are constants equal to the neutral element of $G$ on the
boundary circle.

Writing a general element $f\in \mathcal G$ as $f=gk$ with
$g\in D_0G$ and $k\in K$ we can define the determinant as
\begin{eqnarray}
\text{det}(f) = \text{det}(g) \cdot e^{\TR \log(k)}= 
\text{det}(g) e^{\frac{1}{2\pi} \int_D \tr\, k_1} ,\label{eq5}
\end{eqnarray}
The determinant $ \text{det}(g)$ is defined as 
\begin{eqnarray} 
\log \text{det}(g) = \int_0^1 \text{TR} (g(t)^{-1} * \partial_t
g(t))\, dt,\label{eq6}
\end{eqnarray}
where $g(t)$ (with $0\leq t\leq 1$) is a homotopy in $D_0 G$ joining
the neutral element $g(0)$ to $g=g(1).$ One should remember that the
inverse $g(t)^{-1}$ is defined with respect to the star and matrix
product, so it contains terms of higher order in $\nu.$ 
This determinant for the star product
algebra was introduced by Melrose and Rochon in  \cite{MR}  in connection with a construction of
determinant  line bundles over pseudodifferential  operators.  

The expression $\text{TR} (g^{-1}* dg)$ is a closed form on $\mathcal G$ by the tracial property
of $\TR$ and for this reason $\log \, \det(g)$ depends only on the homotopy class of the 
path $g(t).$  In order that the determinant is well-defined independent of the path one
only needs to check that the integral for generators of $\pi_1(\mathcal G)$ is equal to a 
multiple of $2\pi i:$ 
  
\begin{Theorem}\label{thm1}

Let $G$ be connected and simply connected compact simple matrix Lie group
and $f:S^1 \to \mathcal G$ be a closed smooth loop. Then the 
winding number of the determinant $\text{det}(f(t,\cdot))$ around the loop is equal to the integer
$$\frac{-1}{24\pi^2} \int_{S^1 \times D} \text{tr}\, (f^{-1}df)^3.
$$ 
Here we
have identified the parameter space $D$ as a unit sphere $S^2$ since
on the boundary of $D$ all the functions $f\in\mathcal G$ take the
constant value $1.$ 
\end{Theorem}
{\bf Proof:} The proof is by a direct computation. We need to select a
generator for $\pi_1(\mathcal G)=\mathbb {Z}.$ Since the topology of the group is
determined by the constant part of formal power series in $\nu,$ we
can assume that $f(t,\cdot)$ is zero order in $\nu.$ By the definition
of `TR', we need to compute the term first order in $\nu$ in the
integral (the zeroth order term vanishes identically since $f^{-1} df$
is traceless) 
$$\int_0^1 \text{TR}_{\nu} (f(t,\cdot)^{-1} * f(t,\cdot)) dt.$$
The inverse $f^{-1},$ as defined with respect
to the star product, can be written as   
$$g_0 + \nu g_1 +\nu^2 g_2  + \dots,$$ 
where $g_0$ is the pointwise matrix inverse of the function
$f(t,\cdot)$ and 
$$g_1 = \frac{i}{2}df^{-1} df f^{-1}.$$
Thus 
\begin{eqnarray*}
\int \text{TR} (f^{-1}*\partial_t f)\, dt &=& \frac{1}{2\pi}\int \int_D\frac{i}{2}  \text{tr}\left(
df^{-1}df f^{-1} \partial_t f + df^{-1}d(\partial_t f)\right)\, dt \\ 
&=& \frac{-1}{12\pi i} \int_{[0,1]\times D}  \tr (f^{-1}df)^3
\end{eqnarray*}
which proves the Theorem. 
$\square$

We define 
\begin{eqnarray}
\widehat{LG}= (DG\times S^1)/N, \label{eq7}
\end{eqnarray}
where $N$ is the normal subgroup consisting of pairs $(g,\lambda)$ such that 
$g\in\mathcal G$ and 
$\lambda= \text{det}(g).$ 

This is a central extension by the circle $S^1$ of the loop group
$LG.$ 

\begin{Theorem}\label{thm2}
 The Lie algebra of $\widehat{LG}$ is isomorphic as
a vector space to the direct sum $L\mathbf{g} \oplus i\mathbb{R}$ with the
commutator
$[(f,\alpha),(g,\beta)]=([f,g], c(f,g))$ where $[f,g]$ is the
point-wise commutator of Lie algebra valued functions and $c$ is the
2-cocycle 
\begin{eqnarray}
c(f,g) = \frac{1}{2\pi i} \int_{S^1} \text{tr}\, f dg.\label{eq8}
\end{eqnarray}
\end{Theorem}
{\bf Proof:} Let $\psi$ be the local section of the circle bundle 
$\widehat{LG} \to LG,$  defined in a neighborhood of the unit element
in the loop group, given by 
$$\psi(e^X) = e^{\tilde  X},$$ 
where $X:S^1\to \mathbf{g}$ and $\tilde X\in \mathcal B$  is equal to $X$ on the
boundary. For example, we can fix a smooth function $f(r)$ of the
radius $r$ such that $f(0)=0, f(1)=1$ and all the derivatives of $f$ 
vanish at $r=1$ and put $\tilde X= f(r) X.$  
The exponential is 
defined by the star product,
$$e^Z = \sum_n  \frac{1}{n!} Z*Z*\dots *Z, \text{ $n$ factors}.$$
The section $\psi$ is well-defined in an
open set of $G$ valued of loops where the logarithm is defined.

Locally, near the unit element, the central extension $\widehat{LG}$
is a product of an open set of $LG$ with $S^1.$ The local $S^1$ valued
group cocycle is evaluated from 
\begin{eqnarray}
\text{det} (\psi(e^{\tilde X})\star \psi(e^{\tilde Y}) \star
\psi(e^{-\tilde X})\star \psi( e^{-\tilde Y}) ).\label{eq9}
\end{eqnarray}
The Lie algebra cocycle $c(X,Y)$ is then the bilinear term in the 
expansion of (\ref{eq9}) in powers of $X,Y.$ Using the definition (\ref{eq5}) of the determinant
and the Baker-Campbell-Hausdorff formula 
$$e^X e^Y = e^{X+Y +\frac12[X,Y] + \dots}
$$ 
we obtain
\begin{eqnarray}
c(X,Y) = \text{TR}[\tilde X, \tilde Y]_{\star} = 
\frac{1}{2\pi i} \int_{S^1} \text{tr} X dY.\label{eq10}
\end{eqnarray}
$\square$

The canonical  connection on the loop group $LG$ is given through the $S^1$ invariant 
1-form $\theta$ on $\widehat{LG},$ 
\begin{eqnarray}
\theta = pr_c (g^{-1}dg), \label{eq11}
\end{eqnarray}
where $pr_c$ is the projection onto the center of the Lie algebra
$\widehat{L\gm}.$ 
The curvature form $\Omega$ of this connection is the left invariant 2-form on $LG$ which 
at the identity element is given by the cocycle $c:L\gm\times L\gm \to \mathbb{C}.$ 
The winding number in Theorem \ref{thm1} is then $1/2\pi$ times the integral of $\Omega$ over the 
set  of loops $t\mapsto f(t,x)$ parametrized by $x\in D.$

\section{Generalization to higher dimensions}\label{gensec}

The discussion above cannot directly be generalized to higher dimensions. The obstruction 
is the noninvariance of the boundary conditions. If we have a symplectic manifold with boundary 
of dimension $2d$ then the space of smooth functions with vanishing normal derivatives at the 
boundary is not closed in general. This happens already in the case of a disk in 
$\mathbb{R}^{2d}$ 
with the standard constant symplectic form in $\mathbb{R}^{2d}$. 
For this reason we focus only on a special case. Let $M=D\times S$ where $S$ is a closed manifold 
of dimension $2d-2$ and $D$ is the unit disk in $\mathbb{R}^{2}$. 
We assume that the algebra of functions $\mathcal S$ on $S$ is equipped with a star product  
and $D$ comes with a star product as in Section \ref{disksec}. The star product on
$\mathcal S$ does not need to come 
from a bidifferential operator related to a symplectic form as in the case of the Moyal product.
In fact,  we can
consider as well a product coming from quantum groups or quantum homogeneous spaces. However, what we need
is an 'algebra of functions' possessing a trace functional $\tr_{\mathcal
S}.$   In this case the star product algebra of matrix valued
functions  on $M$ is replaced by the tensor product of the star algebra of
matrix valued functions on the disk and a star algebra  $\mathcal S.$ 
We can now impose  vanishing normal derivatives at the boundary of $D.$  

{\bf Example 1} 

The product of the symplectic disk $D$ and a fuzzy sphere $S^2_N.$ The fuzzy sphere is defined as the quotient 
by an ideal $I$ of the noncommutative associative polynomial algebra in three variables $x,y,z$ with relations 
$x*y -y*x = z, y*z-z*y=x, z*x-x*z=y.$ The two-sided ideal $I$ is generated by the single element $x^2 +y^2 + z^2 + N(N+1)$ 
where $N$ is a nonnegative integer. Since $x,y,z$ define the Lie algebra of $SU(2)$ the trace is defined as the matrix 
trace in an irreducible representation of dimension $N(N+1).$ The algebra is simply the algebra of square matrices 
in dimension $N(N+1).$  

{\bf Example 2} 

We can take as $\mathcal S$ the algebra of smooth
$n\times n$ matrix valued functions in $\mathbb R^{2d-2}$ which decay
faster than any inverse power of $|x|$ at infinity.
The star product is defined as the Moyal product and the trace is the integral of a function 
over $R^{2d-2}.$ In this case the product can actually be defined analytically, not only 
as a formal power series in $\nu.$ This is because the functions can be interpreted as 
symbols of infinitely smoothing pseudodifferential operators in $\mathbb R^{d-1}.$  
This is achieved by
selecting a Lagrangian polarization $\mathbb R^{d-1} \oplus \mathbb R^{d-1}$
and interpreting the first $d-1$ variables as momenta and the last
$d-1$ variables as coordinates. The algebra $\Psi^{-\infty}$ is a subalgebra of
the algebra $\gm_1$ of trace-class operators in the Hilbert space
$H=L^2(\mathbb R^{d-1}, \mathbb C^N).$ 

The linear functional
\begin{eqnarray}
\text{TR}(f)= \frac{1}{2\pi} \int_{D}dx\,dy\, \tr_{\mathcal S} \, f\label{eq12}
\end{eqnarray}
is a trace in the subalgebra of functions which vanish on the
boundary of $D.$  Here $\tr_{\mathcal S}$ denotes the combined 
matrix trace and a trace in in the algebra $\mathcal S.$  
 
The determinants are defined by straight-forward generalization of (\ref{eq5}). 
The Lie algebra cocycle for $Map(S^1, \mathcal S\otimes \gm)$ becomes 
\begin{eqnarray} 
c(f,g)= \frac{1}{2\pi} \int_{S^1} \text{tr}_{\mathcal S} \, f dg.\label{eq13}
\end{eqnarray}

In the case of Example 1 we get the standard central extension of the loop algebra of smooth 
maps from $S^1$ to matrices of size $n N(N+1) \times nN(N+1) $ whereas in the example 2 we have 
a central extension of the loop algebra $L\Psi^{-\infty}$ in the
algebra $\Psi^{-\infty}$ of infinitely smoothing $n\times n$ matrix pseudodifferential operators
over $\mathbb R^{d-1}.$  

The Lie algebra cocycle (\ref{eq13}) extends to the loop algebra $L\gm_1.$ 
A representation for $\widehat{L\gm_1}$ is obtained essentially in the same way as
for central extensions of the loop algebra $L\gm$ based on a finite-dimensional Lie
algebra $\gm.$ 
A highest weight representation of the Lie algebra
$\gm_1$ is given by an infinite increasing sequence of integers
$\lambda_i,$ $i\in\mathbb Z,$  with  $\lambda_i=\lambda_{\infty}$ for $i>>0$ and $\lambda_i=\lambda_{- \infty}\leq \lambda_{\infty}$
for $i<<0.$ 
The irreducible integrable highest weight representation
corresponding to $\lambda$ is then characterized by the existence of a
cyclic vector $v_{\lambda}$ such that 
$$e_{ii}v_{\la} =  \lambda_i  v_{\lambda} \text{ and } 
  e_{ij}v_{\la} =0 \text{ for } i>j,$$ 
where the $e_{ij}$'s are the Weyl basis vectors in $\gm_1,$ 
$$ [e_{ij},e_{kl}] = \delta_{jk} e_{il} -\delta_{il} e_{kj}.$$ 

Given the irreducible highest weight representation $(\lambda)$ of
$\gm_1$ one obtains an irreducible highest weight representation of
the central extension of the loop algebra $L\gm_1$  
by induction. The representation has a highest weight vector
$v_{\la,k}$ characterized by 
$$e_{ij} v_{\la,k}=0 \text{ for } i>j \text{ and } x^{(n)}v_{\la,k}=0 
\text{ for } n<0,$$ 
where $x^{(n)} = e^{in\phi} x \in L\gm_1$ and $k$ is the value of
central element in the representation,
$$c(f,g)= \frac{k}{2\pi} \int_{S^1} \tr(fdg).$$
 The representation integrates to an unitary representation of the
group $\widehat{LG}$ if $k$ is an integer with $\la_{\infty} -\la_{-\infty} \leq k,$ see the monograph by Kac \cite{K}.   

The construction of the central group extension $\widehat{LG}$ for the case of a 
compact matrix group $G$ can now be extended without any changes to the case when $G$ 
is the infinite-dimensional Lie group of unitary pseudodifferential operators $A$ 
such that $A-1$ is trace class.  

{\bf Example 3}

We deform the gauge current algebra in 3 space dimensions. First, let
$\mathbf {n}$ be the ideal of pseudodifferential operators, on a compact spin manifold $M$ of 
dimension 3, of degree less or equal to $-2.$ All pseudodifferential operators are taken with 
matrix coefficients. The matrices act in the tensor product of the spinor bundle and a trivial 
vector bundle $V$ over $M.$ The finite-dimensional Lie algebra $\gm$ of a gauge group $G$ acts 
in the fibers of $V$ through a matrix representation. For each smooth map $X:M \to \gm$ we 
define a deformed operator 
$$\tilde X=  X + \frac{1}{4(D^2 +1)} [D,[D,X]],$$ 
where $D$ is the Dirac operator on $M$ defined by a fixed metric and spin structure. The difference
$\tilde X -X$ is a pseudodifferential operator of order $-1.$ One easily checks that 
$$[\tilde X, \tilde Y] = \widetilde{[X,Y]} \text{ mod } \mathbf {n}.$$ 

Denote by $\mathbf {p}$ the Lie algebra of pseudodifferential operators such that the leading symbols of order 
$0$ and $-1$  are given by the leading symbols of symbols of the deformed operators  $\{\tilde X| X\in Map(M, \gm) \}.$ 
Let $\epsilon= D/|D|.$ Then $[\epsilon,T]$ is Hilbert-Schmidt for all $T\in \mathbf {p}.$  

We have the exact sequence 
$$0\to \mathbf {n} \to \mathbf {p} \to Map(M,\gm),$$ 
where the second map is embedding of Lie algebras and the third map extracts the zero order part
$X$ of an  element $T=\tilde X+ z\in\mathbf {p},$ where $z\in \mathbf {n}.$ 

The Lie algebra $\mathbf {p}$ is a subalgebra of $\mathbf{gl}_{res}$ where the latter consists of bounded 
operators $T$ in the Hilbert space $H$ such that $[\epsilon,T]$ is Hilbert-Schmidt. The algebra 
$\mathbf{gl}_{res}$ has a canonical central extension $\widehat{\mathbf{gl}}_{res}$ defined by the cocycle  
$$c(X,Y)= \frac14 \tr\,\epsilon[\epsilon,X][\epsilon,Y].$$ 
The restriction to $\mathbf {p}$ gives a central extension $\widehat{\mathbf{p}}$ of 
$\mathbf {p}.$ 
Likewise, we have a central extension $\widehat{\mathbf {n}}$ of $\mathbf {n} \subset \mathbf{gl}_{res}.$ 
Putting these together we have the extension
$$0\to \widehat{\mathbf {n}} \to  \widehat{\mathbf {p}} \to Map(M,\gm).$$ 

The algebra $\widehat{\mathbf {p}}$ has unitary highest weight representations. For example, the Fermionic Fock
space $\mathcal F$ based on the polarization $H=H_+\oplus H_-$ carries through canonical quantization a
resepresentation of $\widehat{ \mathbf{gl}}_{res}$ and thus of $\widehat{\mathbf {p}}.$  However, this representation 
does not preserve the domain of the quantization $\widehat D$ of the Dirac operator $D.$


\begin{thebibliography}{99} 

\bibitem{BFFLS} F. Bayen, M. Flato, C. Fronsdal, A. Lichnerowicz, and D. Sternheimer: 
Deformation Theory and Quantization, I--II. {\it Ann. Physics} {\bf 111},  61-151 (1978).

\bibitem{F} B.V. Fedosov:  {\it Deformation Quantization and Index Theory},   Section 5.6.
Akademie Verlag, Berlin, 1996. 

\bibitem{GM} H. Grosse and J. Madore: A noncommutative version of the Schwinger
model. {\it Phys. Lett.} {\bf B283},  218-222 (1992).

\bibitem{GR} S. Gutt and J. Rawnsley: Traces for star products on symplectic 
manifolds.  math.QA/0105089. {\it J. Geom. Phys.} {\bf 42},  No. 1-2, 12-18 (2002).

\bibitem{LVZ} 
F. Lizzi, P. Vitale, A. Zampini: The fuzzy disk. 
{\it J. High Energy Phys.}  0308:057, 16pp. (electronic) (2003). hep-th/0306247.

\bibitem{K} V. Kac: {\it Infinite-dimensional Lie Algebras},  Third Edition,
Cambridge University Press, Cambridge, 1990. 

\bibitem{Ma} J. Madore: {\it Noncommutative Geometry and Applications}, 
Cambridge University Press, Cambridge, 1995.

\bibitem{Mi} J. Mickelsson: Kac-Moody groups, topology of the Dirac
determinant bundle and fermionization. {\it Comm.  Math. Phys.} {\bf 110},
173-183 (1987).

\bibitem{Mi2} J. Mickelsson: {\it Current Algebras and Groups},  Plenum
Press, London and New York, 1989.

\bibitem{MR} R. Melrose and F. Rochon: Periodicity of the determinant line bundle. 
Work in progress.

\bibitem{Mu} M. Murray: Another construction of the central extension of the
loop group. {\it Comm. Math. Phys.} {\bf 116},  73-80  (1988).

\end{thebibliography}
\end{document}